\documentclass[twocolumn,twocolappendix]{aastex631}
\usepackage{graphicx}
\usepackage{grffile}
\usepackage{dcolumn}
\usepackage{xcolor}
\usepackage{color}
\usepackage{bm}
\usepackage{amssymb, amsmath}

\usepackage{natbib}
\usepackage{soul}
\usepackage{times}

\usepackage{cancel}

\begin{document}
	
\preprint{XXX}

%
\title{Turbulence and Magnetic Reconnection in Relativistic Multispecies
	Plasmas}

\author[0000-0002-7459-5735]{Mario Imbrogno} \affiliation{Dipartimento di
	Fisica, Universit\`a della Calabria, Arcavacata di Rende, 87036, Italy}

\author[0000-0001-8694-3058]{Claudio Meringolo} \affiliation{Institut f\"ur
	Theoretische Physik, Goethe Universit\"at, Max-von-Laue-Str. 1, D-60438
	Frankfurt am Main, Germany}

\author[0000-0002-3945-6342]{Alejandro Cruz-Osorio}
\affiliation{Instituto de Astronom\'{\i}a, Universidad Nacional
	Aut\'onoma de M\'exico, AP 70-264, 04510 Ciudad de M\'exico, Mexico}

\author[0000-0002-1330-7103]{Luciano Rezzolla} \affiliation{Institut f\"ur
	Theoretische Physik, Goethe Universit\"at, Max-von-Laue-Str. 1, D-60438
	Frankfurt am Main, Germany} \affiliation{CERN, Theoretical Physics
	Department, 1211 Geneva 23, Switzerland} \affiliation{School of
	Mathematics, Trinity College, Dublin, Ireland}

\author[0000-0001-6295-596X]{Beno\^it Cerutti}
\affiliation{Univ. Grenoble Alpes, CNRS, IPAG, 38000 Grenoble, France}

\author[0000-0002-0786-7307]{Sergio Servidio} \affiliation{Dipartimento
	di Fisica, Universit\`a della Calabria, Arcavacata di Rende, 87036, Italy}




\begin{abstract}
	Simulations of relativistic plasmas traditionally focus on the dynamics
	of two-species mixtures of charged particles under the influence of
	external magnetic fields and those generated by particle
	currents. However, the extreme conditions of astrophysical plasmas near
	compact objects such as black holes and neutron stars are often
	characterized by mixtures of electrons, protons, and positrons, whose
	dynamics can differ significantly because of the considerable mass contrast. We
	present the first two-dimensional particle-in-cell simulations of
	relativistic turbulence and magnetic reconnection in a three-species
	plasma, varying the relative abundance of electrons, protons, and positrons while
	employing realistic mass ratios to achieve unprecedented accuracy. We
	find that turbulence leads to the formation of magnetic islands, current
	sheets, and plasmoids. Reconnection occurs between these structures, with
	plasma composition playing a key role in determining the number of
	reconnection sites and their energy-conversion efficiency. In particular,
	as the proton fraction increases, very small-scale features of the
	turbulence are washed out, while global dissipative effects are amplified.
	Finally, using a novel generalization of Ohm's law for a relativistic
	multi-species plasma, we find that the reconnection rate is primarily
	governed by the electric fields associated to the divergence of the
	positron and electron pressure tensors. These results provide new
	insights into dissipation and particle acceleration in turbulent
	relativistic plasmas, such as those near black holes and neutron stars, and can
	be used to interpret their high-energy emission and phenomenology.
\end{abstract}



\section{Introduction}
\label{sec:intro}

The study of turbulent relativistic plasmas in astrophysical
environments, such as those surrounding black holes and neutron stars, is a crucial
research area that integrates high-energy astrophysics, particle
kinetics, and magnetohydrodynamics (MHD) \citep{balbus1998instability,
	blandford1987particle, biskamp2003magnetohydrodynamic}. These plasmas
may consist of multiple particle species, including electrons, protons,
and positrons, whose concentrations depend sensitively on the specific
astrophysical context. Nevertheless, plasma kinetic effects are essential to
accurately capture global behaviors in low-collisionality environments in
general, and magnetic reconnection in particular.

Although special- and general-relativistic MHD simulations can reproduce a
wide range of reconnection rates in turbulent
plasmas~\citep{shibata2005magnetohydrodynamics, Giacomazzo:2007ti,
	Anderson2008, Giacomazzo:2009mp, etienne2015illinoisgrmhd,
	liska2020large, mosta2013grhydro, Porth2017, ressler2021magnetically},
fluid models are ultimately limited in their ability to describe key
kinetic phenomena, such as nonthermal energy distributions and fast
magnetic reconnection encountered in astrophysical scenarios
\citep{palenzuela2009beyond, yamada2010magnetic, ServidioEA12}. In these
regimes, kinetic effects are expected to play a dominant role in
governing reconnection dynamics.

Arguably, particle-in-cell (PIC) simulations represent the most reliable
ab-initio approach for studying the dynamics of free charges under the
influence of external magnetic fields and those generated by their
currents. However, the vast majority of PIC simulations neglects the
presence of positrons (or other ions) and is restricted to simple
electron-proton or electron-positron (pair) plasmas. While the inclusion
of a third species introduces additional complexity to plasma dynamics,
the presence of a third charge carrier can affect local charge
neutrality, current structures, and the efficiency of energy dissipation
through reconnection~\citep{sironi2014relativistic, Comisso2018}. Despite
their relevance, the dynamics of fully multi-species configurations
remains largely unexplored. Yet, relativistic environments involving
accreting compact objects are expected to host not only electrons and
protons but also considerable portions of positrons, produced by
ultra-high electromagnetic fields accompanying these
environments. Typical scenarios where these conditions are met include
gamma-ray bursts~\citep{piran2004physics, kumar2015physics} and pulsar
magnetospheres~\citep{harding2006physics, philippov2022pulsar}, but also
turbulent phenomena such as magnetic breakouts~\citep{Most2023,
  Kiuchi2023, Musolino2024b}, Kelvin-Helmholtz-like instabilities at the
jet-disk interface, and regions characterized by magnetic reconnection
and plasmoid formation~\citep[see, e.g.,][]{Nathanail2020,
  Nathanail2021b, MellahEA22, vos2024particle}. Improving our
understanding of these environments is essential for modeling key
processes such as jet launching and collimation, large-scale energy
transport, and particle acceleration mechanism responsible for
high-energy emission~\citep{Blandford1977, Blandford1982,
  murray1995accretion, proga2000dynamics, zenitani2001generation,
  mckinney2006general, komissarov2007magnetic, tchekhovskoy2011efficient,
  sironi2015relativistic, Ripperda2020, Camilloni2025}

To overcome these limitations and advance our understanding of particle-field
interactions in relativistic environments, we present the first studies of
relativistic turbulence using
two-dimensional PIC simulations involving three particle species with
realistic mass ratios. By systematically varying the positron-to-electron
concentration ratio in a globally neutral plasma, we analyze the
properties of reconnection regions and show that plasma composition
plays a fundamental role in determining both the number of reconnection sites and
their dissipation efficiency.

\begin{table*}
  \begin{center}
    \begin{tabular}{cccccccccccc}
      \hline
      \hline
      Simulation & $\chi$ & $\beta_{+}$ & $\beta_p$ & $\theta_e$ & $\theta_p$ & $B_{0,z}$ & $PP\lambda_{D,-}^{2}$ & $\lambda_{D,-}/\Delta x$ & $\lambda_{D_{+}}/\Delta x$ & $\overline{\mathcal{R}}_{\text{XP}}$ & $\zeta$ \\
      \hline
      \texttt{chi.0.1} & $0.1$ & $1 \times 10^{-2}$ & $9 \times 10^{-2}$ & $137.79$ & $7.50 \times 10^{-2}$ & $52.50$ & $18585$ & $8.80$ & $27.84$ & $0.050$ & $-13.4$ \\
      \hline
      \texttt{chi.0.4} & $0.4$ & $4 \times 10^{-2}$ & $6 \times 10^{-2}$ & $91.92$ & $5.01 \times 10^{-2}$ & $42.88$ & $12407$ & $7.19$ & $11.37$ & $0.036$ & $-19.4$ \\
      \hline
      \texttt{chi.0.5} & $0.5$ & $5 \times 10^{-2}$ & $5 \times 10^{-2}$ & $76.63$ & $4.17 \times 10^{-2}$ & $39.15$ & $10359$ & $6.57$ & $9.28$ & $0.031$ & $-20.5$ \\
      \hline
      \texttt{chi.0.6} & $0.6$ & $6 \times 10^{-2}$ & $4 \times 10^{-2}$ & $61.33$ & $3.34 \times 10^{-2}$ & $35.02$ & $8269$ & $5.87$ & $7.58$ & $0.030$ & $-22.1$ \\
      \hline
      \texttt{chi.0.9} & $0.9$ & $9 \times 10^{-2}$ & $1 \times 10^{-2}$ & $15.46$ & $8.42 \times 10^{-3}$ & $17.58$ & $2088$ & $2.95$ & $3.11$ & $0.013$ & $-45.6$ \\
      \hline
    \end{tabular}
  \end{center}
  \caption{Summary of the parameters for the five simulations presented
    here. Different simulation setups, labeled as \texttt{chi.0.1},
    \texttt{chi.0.4}, \texttt{chi.0.5}, \texttt{chi.0.6}, and
    \texttt{chi.0.9}, are detailed. From left to right, the columns list:
    the global positron-to-electron concentration ratio $\chi$; the
    plasma beta for positrons ($\beta_{+}$) and protons ($\beta_p$) with
    the electron plasma beta fixed at $\beta_{-} = 0.1$; the
    dimensionless temperature for electrons and positrons, $\theta_e =
    \theta_{-} = \theta_{+}$; the out-of-plane magnetic-field strength,
    $B_{0, z}$; the number of particles per unit area, expressed as
    $PP\lambda_{D,-}^{2}$ per unit Debye area; the ratio of the electron
    and positron Debye lengths to the grid spacing,
    $\lambda_{D,\pm}/\Delta x$; the median reconnection rate (in absolute
    value) at X-points, $\overline{\mathcal{R}}_{\text{XP}}$; the slope
    $\zeta$ of the linear fit to the PDFs of the reconnection rates at
    X-points.}
  \label{TableI}
\end{table*}

\section{Methods}

Simulations are based on a fully kinetic model of relativistic plasma,
implemented using the PIC code \texttt{Zeltron}~\citep{CeruttiEA13},
which solves the equations of motion for a distribution of charged
particles coupled to Maxwell's equations. The former are evolved
using a standard Boris algorithm~\citep{Boris1973}, which is
second-order accurate, symplectic, and preserves phase-space volume.
The latter are expressed in terms of the total magnetic field $\bm{B}$,
the electric field $\bm{E}$, the current density $\bm{J} := \sum_{a}
q_{a} n_{a} \bm{V}_{a}$, and the charge density $\rho_c := \sum_{a} q_{a}
n_{a}$. Here, the index $a$ labels the particle species, so that $n_{a}$
denotes the number density of species $a$ (we use ``$p$'' for protons,
``$+$'' for positrons, and ``$-$'' for electrons), while $\bm{V}_{a}$ and
$q_{a}$ represent the corresponding bulk velocity and charge. Taking
$q_{-}$ and $m_{-}$ as the electron charge and mass, respectively, we set
$q_{-} = -q_{+} = -q_p = -e$ and $m_{-} = m_{+}$, where $e$ is the
elementary charge. Under the (reasonable) assumption of global plasma
neutrality, the species number densities must satisfy $n_{-} = n_{+} +
n_p =: n_0 = (4 \pi)^{-1}$. We vary the composition of the three-species
plasma through the positron-to-electron concentration ratio, defined as
$\chi := {n_{+}}/{n_{-}}$, which enable us to explore the meaningful
physical range
\begin{equation}
	\chi =\left\{
	\begin{array}{cll}
		0   & \qquad \text{if} \quad n_{+} = 0\,;    & n_{-} = n_{p}\,,\\
		1/2 & \qquad \text{if} \quad n_{+} = n_{p}\,; & n_{-} = 2n_{p} = 2n_{+}\,,\\
		1   & \qquad \text{if} \quad n_{p} = 0\,;    &  n_{-} = n_{+}\,
	\end{array}
	\right.
\end{equation}
where $0 \leq \chi \leq 1$.

The simulation setup follows that discussed
by~\citet{meringolo2023microphysical}, using a computational grid of $N_x
= N_y = 8192$ mesh points in a square domain of side length $L_0 \approx
10923 \, d_{-}$, where $d_{-} := c/\omega_{p,-}$ is the electron skin
depth, $\omega_{p,-} := \sqrt{4 \pi \, n_0 \, e^2/m_-}$ the electron
plasma frequency, and $c$ the speed of light\footnote{Adopting units
	where $c = 1 = e = m_- = k_{_{\rm B}}$ implies that $d_{-} = 1$ as well,
	where $k_{_{\rm B}}$ is the Boltzmann constant.}.
Importantly, to avoid artifacts and systematic biases, we employ a
realistic proton-to-electron mass ratio of $m_p/m_- =
1836$~\citep[see][for a discussion on the importance of employing
realistic mass ratios]{Rowan2017}, and a total of approximately $1.6
\times 10^{10}$ macroparticles, thereby minimizing noise in both the
fields and particle moments.

To accurately simulate multi-species plasmas, it is important to adopt
spatial resolutions that capture key physical scales, such as the Debye
length and the inverse plasma frequency, while ensuring numerical
convergence. At the same time, an optimal balance between computational
efficiency and physical accuracy must be maintained. If $\lambda_{D,-} :=
\sqrt{m_-\, \theta_{-}\, c^2 / 4 \pi \, n_0 \, e^2}$ denotes the electron
Debye length, with $\theta_{-}$ the dimensionless electron temperature,
then the spatial and temporal resolutions must be chosen such that
$\Delta x < \lambda_{D,-}$ and $\Delta t < (\omega_{p,-})^{-1}$. We set
$\Delta x = 4\, d_{-}/3 \sim (0.11 \!-\! 0.34) \, \lambda_{D,-}$ (see
Tab.~\ref{TableI}) and $\Delta t = 0.45 \, \Delta x = 0.6 \,
(\omega_{p,-})^{-1}$ (see Appendix~\ref{sec:consistency} for a
consistency study). We recall that when particles become relativistic,
their skin depth increases proportionally to the square root of the
Lorentz factor. Specifically, $d_a^{\rm rel.} = \sqrt{ (\gamma_a m_a
	c^2)/(4 \pi n_a e^2) } = \sqrt{ \gamma_a }\, d_a^{\rm nonrel.}$,
where $\gamma_a$ denotes the Lorentz factor. Because the average
Lorentz factors are $\langle\gamma_-\rangle \sim \langle\gamma_+\rangle
\approx 10 - 10^2$, our resolution is effectively higher than simple
non-relativistic considerations may suggest. Moreover, the use of $240$ 
particles per cell (PPC) effectively mitigates noise and guarantees 
satisfactory accuracy.

\subsection{Generalized Ohm's Law}

The interpretation of the simulation results requires a proper
understanding of how Ohm's law needs to be modified to account for a
system of coupled, charged multi-fluids. In the context of a relativistic
electron-positron plasma, such a generalization has been considered, for
instance, by~\citet{hesse2007dissipation}, who examined the role of
kinetic dissipation in relativistic magnetic reconnection. In essence, in
a relativistic multi-species plasma, Ohm's law, which relates the
electric field, plasma currents, anisotropic pressure contributions, and
higher-order moments of the distribution function, can be generalized by
considering the second moment of the relativistic Vlasov equation and
treating separately the terms expressing thermal and kinetic
effects. Following a similar methodology, we rewrite the second-order
moment of the Vlasov equation (i.e., the momentum-conservation equation)
for a given species $a$ as (see Appendix~\ref{AppA} for details):
\begin{equation}
	\partial_t (n_{a} m_{a} \bm{U}_{a}) + \bm{\nabla} \cdot
	\bm{{\bm{\Pi}}}_{a} - q_{a} n_{a} (\bm{E} +
	\bm{V}_{a} \times \bm{B}) = 0\,,
	\label{eq:one}
\end{equation}
where $\bm{U}_{a}$ is the bulk four-velocity defined as the
density-weighted average of the corresponding four-velocity and
representing the first moment of the distribution function, and
$\bm{\Pi}_{a}$ is the generalized pressure tensor, which accounts for
both thermal and bulk contributions. More specifically, we decompose this
tensor by distinguishing the purely thermal contributions, associated
with the pressure tensor $\bm{P}_{a}$, from the kinetic ones as
\begin{equation} 
	\bm{{\bm{\Pi}}}_{a} =
	\bm{{P}}_{a} + m_{a} n_{a} \bm{V}_{a} \bm{U}_{a} = \int
	d^3u \, m_{a} \frac{\bm{u} \bm{u}}{\gamma} f_{a}\,.
\end{equation}
In the expression above, $f_{a} = f_{a} (\bm{x}, \bm{u}, t)$
represents the particle distribution function, and $\bm{u} :=
\gamma \bm{v}$ denotes the four-velocity of the particles, with
$\bm{v}$ and $\gamma$ being the three-velocity and Lorentz
factor, respectively. Note that the compact form of the
momentum-conservation equation~\eqref{eq:one} follows from using the
continuity equation to cancel the divergence of convective inertia term
$\bm{\nabla} \cdot (m_a n_a \bm{V}_a \bm{U}_a)$ with the corresponding
contribution arising from the divergence of the pressure tensor
$\bm{P}_a$.

After combining the contributions from all three species considered, the
resulting expression for the electric field defines the generalized Ohm's
law for a three-species relativistic plasma, given by:
\begin{equation}
	\begin{aligned}
		\bm{E} = & \frac{m_-}{e^2 \mathcal{N}} \, \partial_t \bm{\mathcal{J}} -
		\frac{1}{\mathcal{N}} \left( n_{+} \bm{V}_{+} + n_{-} \bm{V}_{-} \right)
		\times \bm{B} \\ \phantom{=} & - \frac{1}{\mathcal{N}} \, n_p \bm{V}_p
		\times \bm{B} - \frac{1}{e \, \mathcal{N}} \, \bm{\nabla} \cdot
		\bm{{\bm{\Pi}}}_{-} \\ \phantom{=} & +
		\frac{1}{e \, \mathcal{N}} \, \bm{\nabla} \cdot
		\bm{{\bm{\Pi}}}_{+} + \frac{m_-}{m_p}
		\frac{1}{e \, \mathcal{N}} \, \bm{\nabla} \cdot
		\bm{{\bm{\Pi}}}_p \\ = & \bm{E}_{\partial_t} +
		\bm{E}_{vb \pm} + \bm{E}_{vbp} + \bm{E}_{\bm{\Pi} -} +
		\bm{E}_{\bm{\Pi} +} + \bm{E}_{\bm{\Pi} p}\,,
	\end{aligned}
	\label{eq:Ohm}
\end{equation}
where $\mathcal{N} := n_p {m_-}/{m_p} + n_{+} + n_{-}$ is the total effective
number density, $\bm{\mathcal{J}} := \sum_{a} q_{a} n_{a} \bm{U}_{a}$
represents the current density associated with the bulk four-velocity,
and the different terms in the last line of Eq.~\eqref{eq:Ohm} correspond
to those appearing on the right-hand side.

Expression~\eqref{eq:Ohm} offers a concise yet comprehensive description of
the six different contributions to the electric field in a
relativistic three-species plasma. To the best of our knowledge, this is
the first time that an expression for a relativistic three-species Ohm's
law has been presented. As we discuss below, each term can
be monitored and characterized to support the interpretation of our
simulation results.

\begin{figure*}
  \centering
  \includegraphics[width=2.10\columnwidth]{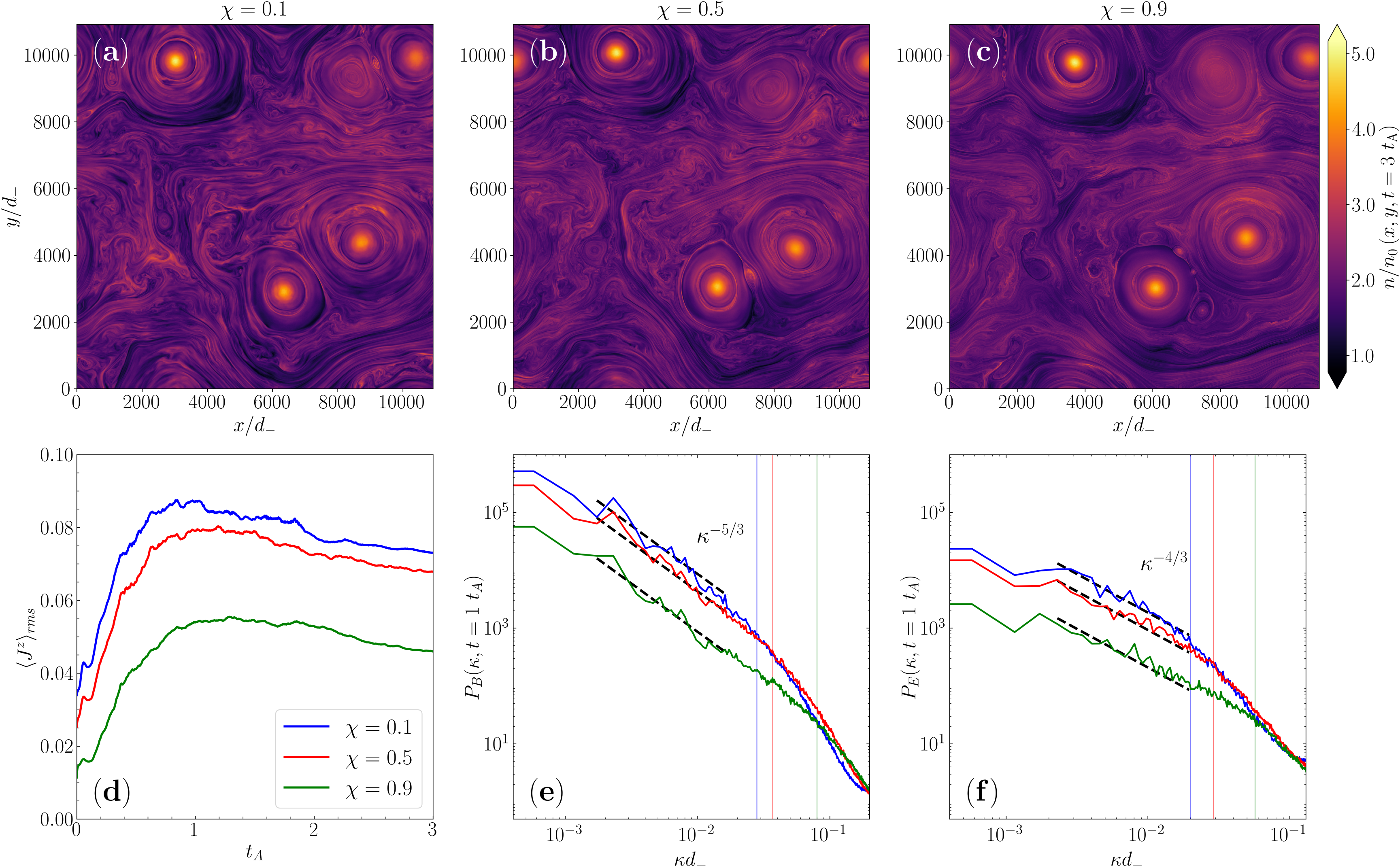}
  \caption{\textit{Upper row:} Snapshots of the (normalized) total number
    density $n/\langle n \rangle_{\mathrm{rms}}$ at time $t = 3 \,
    t_{\text{A}}$ for $\chi = 0.1$ (a), $\chi = 0.5$ (b), and $\chi =
    0.9$ (c). \textit{Lower row:} Panel (d) shows the evolution of the
    rms of the out-of-plane component of the current density for
    different concentration ratios; (e) shows the power spectrum of the
    magnetic field when turbulence is fully developed at $t = 3\,
    t_{\text{A}}$, with the corresponding slopes; (f) shows the same as
    in (e), but for the electric field. The power spectra are normalized
    to the rms of the in-plane magnetic field. Thin solid lines mark the
    spectral slope breaks in the inertial range.}
  \label{Image_1}
\end{figure*}

\subsection{Initial Conditions}

Assuming a Cartesian coordinate system where the plasma dynamics unfolds
in the $(x,y)$ plane, we establish a strongly turbulent regime by
initializing a large-scale, random magnetic field determined via a
superposition of low-wavenumber Fourier
modes~\citep{meringolo2023microphysical, Meringolo2024}, combined with
a uniform background magnetic field directed along the $z$-direction
$B_{0,z}$. More specifically, the initial magnetic field
configuration is such that $\langle B_{xy} \rangle_\text{rms}/B_{0,z} \sim
1$~\citep{ServidioEA12}, where $\langle \delta B_{xy} \rangle_\text{rms}$
represents the root-mean-square (rms) fluctuations of the in-plane
magnetic field.

After defining the total enthalpy density as $\rho h = \rho + \mathcal{U}
+ p$, where $\rho := \sum_{a} n_{a} m_{a} c^2$ is the (total) rest-mass
density, $\mathcal{U} \approx \sum_{a} p_{a} (\Gamma_{a} - 1)^{-1}$ the
internal energy density, and $p := \sum_{a} p_{a}$ the total
pressure~\citep{Rezzolla_book:2013}, we introduce the magnetization
parameter $\sigma := p_{\rm mag}/\rho h = B^2/\rho h$ to distinguish
between regions that are magnetically dominated (i.e., $\sigma \gg 1$)
and those that are not (i.e., $\sigma \lesssim 1$). A couple of remarks
should be made. First, alternative definitions of magnetization exist
that involve only the rest-mass density~\citep[see,
e.g.,][]{Porth2019}. Second, although our plasma is
collisionless, we model it using an ideal-gas equation of
state~\citep{Rezzolla_book:2013} with an adiabatic index $\Gamma_a$ for
each species; as customary, we set $\Gamma_a = 4/3$ for all
species~\citep{ball2018electron}. Similarly, we introduce the total
plasma beta parameter, $\beta := \sum_a \beta_a$, as the ratio of plasma
to magnetic pressure, where $\beta_{a} := 8 \pi n_{a} k_{_{\rm B}} T_{a}
/ B_{0,z}^2 = 8 \pi n_{a} \theta_{a} m_{a} c^2 / B_{0,z}^2$, and $T_{a}$
is the effective temperature of each species.

We note that, despite the numerous degrees of freedom in our setup, it is
possible to construct a controlled set of simulations in which the role
of the concentration ratio $\chi$ can be isolated by keeping $\beta$,
$\sigma$, and the turbulence levels $\langle B_{xy}
\rangle_\text{rms}/B_{0,z}$ fixed. In this way, we simulate five
different plasma configurations, each regulated by different values of
$\chi$ (see Tab.~\ref{TableI} for a summary of the key properties),
explore the dynamics of a three-species trans-relativistic plasma, and
assess the implications for relativistic astrophysical plasmas around
compact objects\footnote{Due to the considerable mass difference, $B_0$
needs to be adjusted for different values of $\chi$ to preserve constant
values of $\sigma$ and $\beta$.}.

Since $\sigma$ and $\beta$ determine the main properties of the plasma
and its dynamics, our initial conditions for the magnetization and plasma
beta are inspired from GRMHD simulations of accretion flows
onto black holes~\citep{Porth2019, Nathanail2020, Ripperda2020,
	Nathanail2021b, meringolo2023microphysical}. Hence, we set $\beta
\approx 0.2$ (with $\beta_{-} \approx 0.1$) and $\sigma \approx 1$. This
choice ensures a balance between magnetic and kinetic forces,
facilitating characteristic plasma behaviors such as vortex formation and
magnetic stability, while still allowing for instabilities to develop. In
this weakly relativistic regime, often referred to as
``trans-relativistic'', protons remain non-relativistic, while electrons
and positrons exhibit a relativistic behavior~\citep[see][for a
proton-electron plasma in this regime]{ball2018electron}.

Hereafter, and for compactness, we focus on a detailed discussion of three
representative configurations (i.e., $\chi = 0.1, 0.5$, and $0.9$), with
the results from the other setups falling in between those presented.

\section{Results}

\subsection{Turbulence and Dissipation}

Given the chosen initial conditions, turbulence develops very rapidly and
saturates after approximately one Alfv\'en time $t_{\text{A}} :=
L_0/v_{\text{A}}$, where $v_{\text{A}} := c \sqrt{\sigma/(1+\sigma)}$ is
the Alfv\'en speed. After this time, the system reaches a quasi-steady
state, as indicated by the saturation of the rms of the out-of-plane
($z$) component of the total current, $J^z$, as shown in
Fig.~\ref{Image_1}-(d). Once turbulence is fully developed, magnetic
islands begin to merge - regions of confined magnetic flux (see
also below) - leading to an inverse cascade of magnetic helicity and
the generation of long-lived, large-scale
structures~\citep{imbrogno2024long}. These large-scale islands are
illustrated in panels (a)--(c) of Fig.~\ref{Image_1}, which show the
total number density at $t = 3\,t_{\text{A}}$. Note that as $\chi$
decreases (and, consequently, the proton number density increases), very small-scale
features are washed out (i.e., only the largest anisotropies
persist) and dissipative effects become more pronounced. This response
is quantified by the total rms out-of-plane current $\langle J^z
\rangle_\text{rms}$ [Fig.~\ref{Image_1}-(d)], which is smaller for
increasing values of $\chi$. A similar trend can also be inferred in
terms of plasmoid formation. In particular, as $\chi \to 1$,
corresponding to a pure pair-plasma configuration, the elongated current
sheets between magnetic islands become unstable and break up at what we
refer to as ``X-points'' (see below for a definition), forming sequences
of plasmoids that are subsequently
accelerated~\citep{bessho2005collisionless, sironi2016plasmoids} [see
Fig.~\ref{Image_1}-(c)].

Panels (e) and (f) of Fig.~\ref{Image_1} report the spectra of the
magnetic and electric fields at $t = 3\,t_{\text{A}}$ for the three
representative mixtures. The slopes observed over approximately one
decade in normalized wavenumber, $\kappa\,d_{-}$, indicate that the
magnetic power spectrum is steeper, scaling as $\propto \kappa^{-5/3}$,
compared to the electric power spectrum, which scales as $\propto
\kappa^{-4/3}$. While the former scaling matches the expectations
for fully developed isotropic turbulence, the latter likely contains
additional electric-field contributions arising from anisotropic
stresses beyond the ideal-MHD approximation. Similar scalings have also
been reported in non-relativistic
turbulence~\citep{gonzalez2019turbulent} and magnetic
reconnection~\citep{adhikari2023effect, lewis2023magnetospheric}.
Notably, these behaviors align with several studies of solar-wind
turbulence~\citep{BaleEA05, AlexandrovaEA09, SahraouiEA09}. It is also
worth remarking that the scales at which the turbulent cascades are cut
off vary with the mixture composition. Specifically, at these
resolutions, the cut-off shifts to smaller scales, down to $\kappa d_{-}
\lesssim 0.08$ when $\chi \to 1$, while larger scales, up to $\kappa
d_{-} \lesssim 0.02$, are suppressed when $\chi \to 0$ (see colored
vertical lines).  The loss of power at large $\kappa$'s begins earlier
for $\chi=0.1$ than for $\chi=0.9$, indicating that the presence of
protons is effectively associated with the ``washing out'' of turbulence
at the smallest scales.

\subsection{X-points, Electric fields, and Reconnection Rates}

\begin{figure}
  \centering
  \includegraphics[width=0.99\columnwidth]{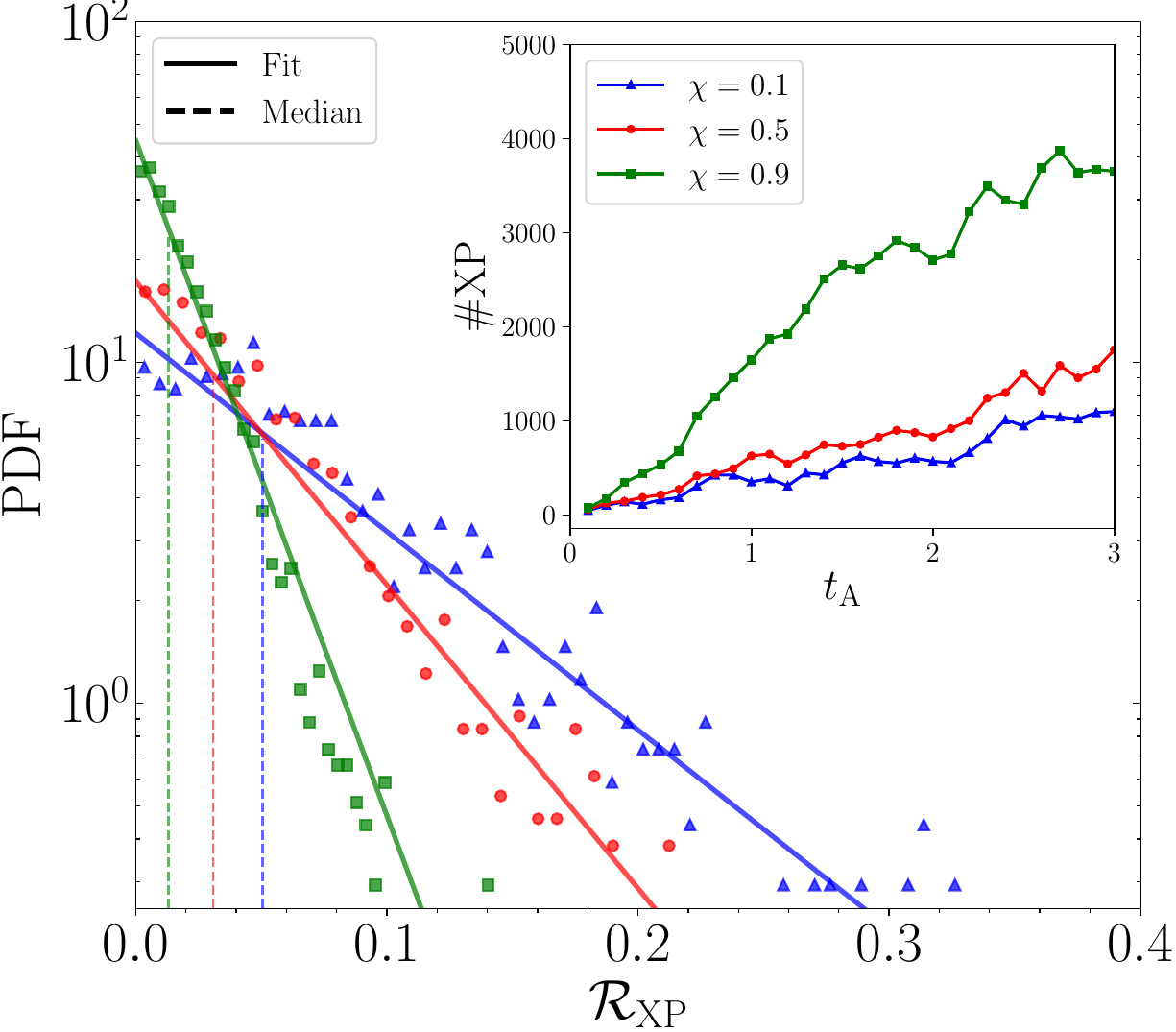}
  \caption{PDFs of the reconnection rates (symbols) for $\chi = 0.1$
    (blue), $\chi = 0.5$ (red), and $\chi = 0.9$ (green), with
    corresponding linear fits shown as solid lines in matching
    colors. Dashed vertical lines indicate the median reconnection rates
    for each distribution, while the inset shows the evolution of the
    number of X-points (${\rm \#XP}$) for each configuration.}
  \label{Image_3}
\end{figure}

\begin{figure}
  \centering
  \includegraphics[width=0.99\columnwidth]{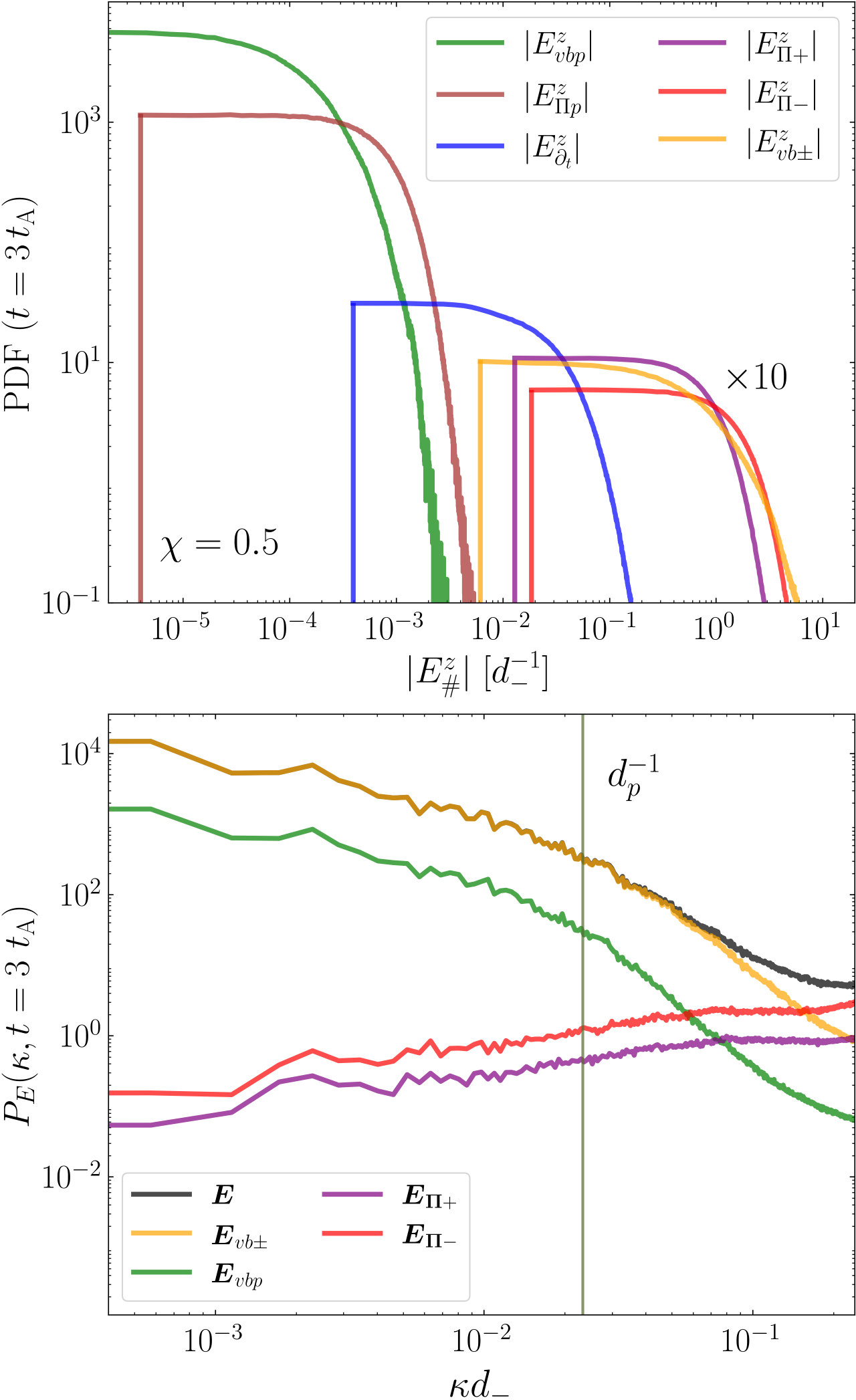}
  \caption{\textit{Top panel:} PDFs of the various contributions to the
    electric field in the $z$-direction, as given in Eq.~\eqref{eq:Ohm},
    for $\chi = 0.5$ and at $t = 3\,t_\text{A}$. \textit{Bottom panel:}
    Power spectra of the dominant contributions reported in the top
    panel. Marked with the gray vertical line is the typical proton
    scale.}
  \label{Image_2}
\end{figure}

Because X-points play a particularly important role in astrophysics, we
have focused on their occurrence in our simulations and the analysis of
various physical quantities at these locations. In particular, at any
given time, we compute the Hessian matrix of the $z$-component of the
vector potential, $A^z(x,y)$, defined as $H^{A^z}_{i,j}(\bm{x}) :=
\partial^2 A^z/\partial x_i \partial x_j$. At each neutral point, where
$\bm{\nabla} A^z \times \bm{e}_z = 0$, we evaluate the local eigenvalues
and eigenvectors of $H^{A^z}_{i,j}$~\citep[see][for details on the
  implementation in the presence of a stochastic
  field]{servidio2010statistics}. We then classify the neutral point as
an X-point if it is a saddle point, or as a magnetic island if it
corresponds to an extremum~\citep{servidio2009magnetic}. Using this
technique, we identify $\sim 10^2$ X-points in the very early stages of
the simulation, with their number increasing over time and as a function
of $\chi$ (see inset in Fig.~\ref{Image_3}).

Having isolated X-points and magnetic islands, we were able to separately
assess the role of the different contributions to the electric field in a
three-species plasma, as described by Eq.~\eqref{eq:Ohm}. The
corresponding results are shown in the top panel of Fig.~\ref{Image_2},
which displays the probability distribution function (PDF) of the various
terms contributing to the $z$-component of the electric field at $t = 3
\,t_{\text{A}}$, computed over the entire domain for the case with $\chi
= 0.5$ (see Fig.~\ref{Image_2_appendix} in Appendix~\ref{AppB} for the
behavior at different concentrations). Note that some components have
been amplified by a factor of ten to be visible on the same scale. A
quick inspection of Fig.~\ref{Image_2} reveals that the largest
electric-field contribution comes from what we refer to as the
``proton-induced electromotive-force'' term, $E^z_{vbp}$ (green line),
followed by the proton pressure-gradient term, ${E}^z_{{\Pi}p}$ (brown
line). Interestingly, a sizeable contribution is also provided by the
displacement term of the momentum current density, $E_{\partial_t}^z$
(blue line), consistent with the findings of~\citet{hesse2007dissipation}
in the presence of a guide magnetic field. Note that although these
electric fields are quite common, they are also very weak and at least
three orders of magnitude weaker than the other electric-field
contributions, such as $E^z_{{\Pi} -}$, $E^z_{{\Pi} +}$, and
$E^z_{vb\pm}$, which are four orders of magnitude less common.

The bottom panel of Fig.~\ref{Image_2} provides the spectral distribution
of the various components of the electric field, revealing that at
small scales the component $\bm{E}_{vb\pm}$ becomes less dominant,
despite being primarily responsible for the overall large-scale
behavior of the total electric field. In contrast, the components
$\bm{E}_{\bm{\Pi}-}$ and $\bm{E}_{\bm{\Pi}+}$ contribute most
significantly at small scales. In fact, at the smallest scales resolved
by our simulation and for $\chi = 0.5$, we find $P_{\bm{E}_{\bm{\Pi}-}}
\sim P_{\bm{E}_{vb\pm}}$. Overall, Fig.~\ref{Image_2} presents a
comprehensive and intriguing picture of the electric-field distribution,
highlighting a landscape where weak but frequent electric fields coexist
with strong yet rare ones. The implications of this complex structure
will be explored in detail in a separate work~\citep{meringolo2025}.

Since X-points coincide with reconnection sites, we use our
time-dependent map of their location and distribution to
calculate the reconnection rate as
\begin{equation}
	\langle \mathcal{R}_{\text{XP}} \rangle := \langle
	|E^z(\text{XP})/\langle B_{xy} \rangle_\text{rms}|
	\rangle\,,
	\label{eq:RXP_Ez}
\end{equation}
namely, the average of the ratio of the $z$-component of the electric
field at the X-point to the rms of the in-plane magnetic field,
expressed in units of the speed of light \citep{zenitani2001generation,
	kagan2015relativistic, liu2015scaling, uzdensky2016magnetic}.

Figure~\ref{Image_3} illustrates the PDFs of the reconnection rates for
all of the X-points across the whole computational domain and the three
reference cases, along with their respective linear fits. Although the
data exhibit significant variance, they suggest that lower values of
$\chi$ leads to higher median reconnection rates (vertical dashed
lines). Furthermore, the median reconnection rate demonstrates a clear
linear dependence on the mixture composition, which can be approximated
as $\overline{\mathcal{R}}_{\text{XP}} \approx a_1(1 - \chi)$, with $a_1
\approx 0.06 \pm 0.0004$ (see also Tab.~\ref{TableI}). The inset in
Fig.~\ref{Image_3} displays the time evolution of the number of X-points
for the three plasma species. Interestingly, the nearly pure pair plasma
($\chi = 0.9$) produces almost three times as many X-points as plasmas
containing a fraction of protons, with their number continuing to
increase over time. In contrast, the nearly pure electron-proton plasma
($\chi = 0.1$) yields the lowest number of X-points, with the latter
remaining relatively stable throughout the evolution. At the same time,
the $\chi = 0.1$ plasma also leads to extreme reconnection processes with
$\mathcal{R}_{\text{XP}} \simeq 0.4$, which, albeit rare, may be
responsible for the most interesting astrophysical phenomenology such as
coronal mass ejections.
We should also note that the inclusion of positrons significantly alters
the dynamics of classical proton-electron plasmas, disrupting the current
layers and weakening local reconnection processes and dramatically
changing the reconnection rate. Hence, while in the limit $\chi \to 1$
protons are absent, the electron-positron plasma that is produced in this
case is very different from a pure-electron plasma
\citep{phan2018electron, sharma2019transition}. The latter is much closer
to an idealized plasma and thus exhibits a rather large reconnection
rate.
Additional key quantities related to the reconnection rates as a function
of $\chi$, including the averages of their absolute values and the slopes
of the linear fits shown in Fig.~\ref{Image_3}, are summarized in
Tab.~\ref{TableI}.

\begin{figure*}[t]
\centering
\includegraphics[width=\textwidth]{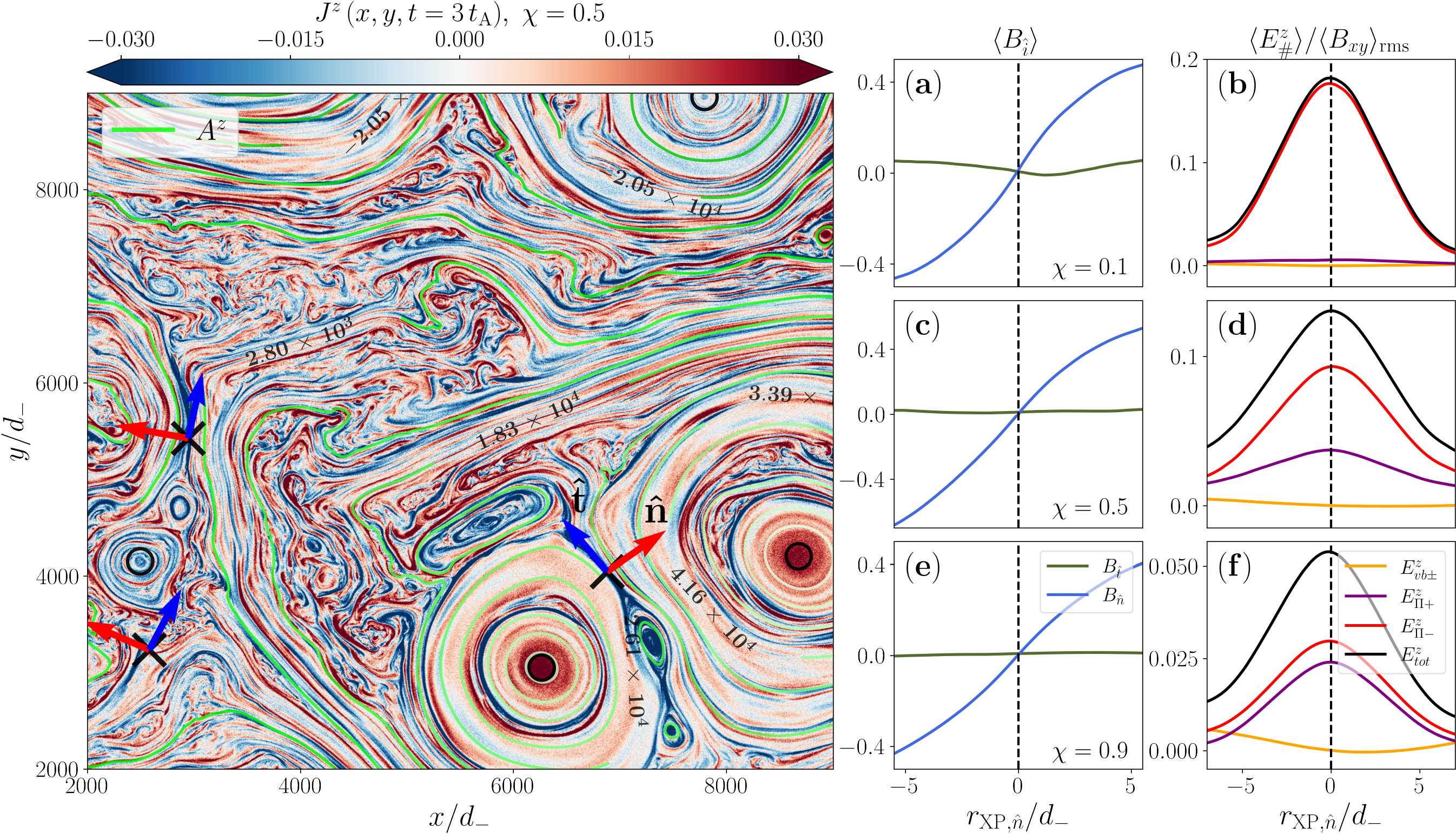}
\caption{\textit{Left panel:} colormap of the current density in the
$z$-direction at $t = 3 \, t_{\text{A}}$ for the simulation with
$\chi = 0.5$. Bright green lines indicate isocontours of the
out-of-plane component of the vector potential $A^z$. Several strong
X-points are marked with an $\bm{\times}$ and serve as origin for the
normal and tangent unit vectors $\bm{\hat{n}}$ (red arrows) and
$\bm{\hat{t}}$ (blue arrows), respectively. \textit{Right panels:}
the left column shows the average profiles of the magnetic-field
components parallel (blue lines) and perpendicular (gray lines) to
the current sheet, plotted along the $n$-direction, for $\chi = 0.1$
(top row), $\chi = 0.5$ (middle row), and $\chi = 0.9$ (bottom
row). The right column displays the average profiles of the various
contributions to Ohm's law, including the total electric field (black
lines), the positron pressure-gradient contribution (purple lines),
the electron pressure-gradient contribution (red lines), and the
electron-positron electromotive contribution (orange lines), also
along the $n$-direction. As before, each row correspond to a
different value of $\chi$.}
\label{Image_4}
\end{figure*}

In addition to the global statistical overview presented in
Fig.~\ref{Image_3}, we also investigate the behavior of the
electric and magnetic fields at X-points. To gain insight into the most
intense reconnection sites, we consider the ``strongest'' X-points,
namely those for which the reconnection rate exceeds a threshold,
$\mathcal{R}_{\text{XP}} > \mathcal{R}_{\text{XP,th}}$, where
$\mathcal{R}_{\text{XP,th}}$ is defined as one standard deviation above
the median. For the cases with $\chi = 0.1, \, 0.5$ and $0.9$, the corresponding
thresholds are $\mathcal{R}_{\text{XP,th}} = 0.12, \, 0.08,$ and $0.03$,
respectively. These locations, representing $ \approx 18 \, \%$ of
the total X-points, are analyzed by evaluating one-dimensional averages
of the magnetic field components tangent ($\bm{\hat{t}}$; blue arrows in
the left panel of Fig.~\ref{Image_4}) and orthogonal to the current sheet
($\bm{\hat{n}}$; red arrows), i.e., $B_{\hat{n}}$ and $B_{\hat{t}}$,
respectively [panels (a), (c), and (e) of Fig.~\ref{Image_4}], as well as
the electric-field components $E^z_{{\Pi} -}$ and $E^z_{{\Pi} +}$,
noticing that the convective electric field component $E^z_{vb \pm}$ vanishes
at any X-point [panels (b), (d), and (f) of Fig.~\ref{Image_4}].

As shown in panels (a), (c), and (e) of Fig.~\ref{Image_4}, the magnetic field
component perpendicular to the current sheet ($B_{\hat{n}}$)
changes sign at the X-point, while the tangential component
($B_{\hat{t}}$) remains zero, in agreement with the Sweet-Parker
model~\citep{parker1957sweet}. Similarly, panels (b), (d), and (f)
illustrate that all the averaged electric field contributions peak at the
X-point, with the electron pressure-gradient term $E^z_{\Pi -}$ being the
dominant one. As $\chi$ increases, the electric field
associated with the positron pressure-gradient, $E^z_{\Pi +}$, also grows,
becoming comparable to $E^z_{\Pi -}$ when $\chi = 0.9$. These two terms emerge
as the dominant contributions to the electric field in these regions,
directly governing the reconnection process and acting as the main
drivers of particle acceleration.

This behavior is not difficult to interpret: when $\chi = 1$ (i.e., when
no protons are present), the electric fields arising from the pressure
gradients of electrons and positrons contribute equally. However, when
$\chi < 1$ (i.e., when protons are present), the electric fields from
positron pressure-gradients compete with those from protons and become
less dominant, causing differences in acceleration between electrons and
positrons to emerge.

To quantify how the electric fields related to pressure-gradient terms differ
as $\chi$ changes, we computed the ratio $E^z_{{\Pi}+} / E^z_{{\Pi}-}$ at
the X-points for different values of $\chi$. We found that this ratio can be
approximated by the fitting expression $E^z_{{\Pi}+}/E^z_{{\Pi}-} \approx
0.4 \chi^2 + 0.6 \chi$. Clearly, the ratio $E^z_{{\Pi}+}/E^z_{{\Pi}-} \to
0$ when no positrons are present $(\chi \to 0)$, while
$E^z_{{\Pi}+}/E^z_{{\Pi}-} \to 1$ for a pure pair plasma ($\chi
\to 1$).

\section{Conclusions}
\label{sec:conclusions}

Using accurate PIC simulations, we have conducted a systematic analysis
of turbulent relativistic plasmas comprising three species of charged
particles: electrons, positrons, and protons with a realistic mass
ratio. This study represents a significant step towards modeling
astrophysical plasmas around compact objects, where all three species are
expected to coexist. By systematically varying the
positron-to-electron concentration ratio, $\chi$, we have progressively
explored how the presence of positrons alters not only the spectral
properties of turbulent plasmas but also how the relative concentrations of the
three species affect the formation of magnetic islands, current
sheets, and plasmoids.

More specifically, by performing five long-term simulations with a
varying mixture from $\chi = 0.1$ (corresponding to a vanishing positron
fraction) to $\chi = 0.9$ (corresponding to a vanishing proton fraction),
we have highlighted that as $\chi$ increases, smaller-scale features in
the turbulence are gradually washed out (i.e., only the largest
anisotropies persist) and global dissipative effects (i.e., across all
scales) are increased. This behavior can be quantified in three different
ways. First, by analyzing the the total rms out-of-plane current $\langle
J^z \rangle_\text{rms}$, which decreases as $\chi$ increases. Second, by
examining the spectral properties of the turbulent electric and magnetic
fields, which, for small values of $\chi$, exhibit significant higher
powers at small scales and an earlier departure from a power-law
scaling. Lastly, by considering the PDFs of the reconnection rate
measured at the X-points, which clearly show a decrease in median values
as $\chi$ increases.

Overall, these systematic variations in the turbulent and dissipative
phenomenology resulting from the changes in charge composition suggest
that the presence of more protons leads to less frequent but more energetic
reconnection events, with larger reconnection rates. In turn,
dissipation occurs primarily at the smallest scales, disrupting the clear
power-law scaling, while the largest turbulent structures remain
unaffected. By contrast, pure electron-positron plasmas are characterized
by more frequent but less energetic reconnection events, leading to
comparatively lower reconnection rates. As a result, dissipation takes
place at all scales, preserving a clear power-law scaling.

A considerable effort has also been dedicated to defining the
generalization of Ohm's law for a three-species plasma, which provides a
relationship between the electric field, plasma currents, anisotropic
pressure contributions, and higher-order moments of the distribution
function. In this way, it was possible to disentangle the large-scale
contributions of the electromotive force from the small-scale
contributions associated with the gradients of the pressure tensor. More
specifically, at the X-points, where reconnection occurs, the
electromotive terms vanish, and the only contributions to the electric
field arise from the pressure gradients, primarily those from electrons
and positrons (the pressure-gradient term associated with protons is
negligible because of their larger mass). As a result, for large values
of $\chi$, the pressure contributions from electrons and positrons are
comparable, and the reconnection is mediated by multi-plasmoids, leading
to weaker reconnection rates. By contrast, for small values of $\chi$,
the electron pressure term is dominant and small-scale plasmoids are less
frequent, resulting in more energetic reconnection events.

While a full comprehension of this behavior will require additional
studies, a possible explanation is as follows. In the limit $\chi \to 0$,
i.e., for an equal mixture of electrons and protons, the substantial mass
difference between the two will lead to different currents in the two
species and to a smaller total current. Within a turbulent reconnection
scenario, this will translate into a small number of X-points, but also
into a larger reconnection rate because these X-points will have a
smaller number density and can generate a larger dissipation. On the
other hand, in the limit $\chi \to 1$, i.e., for an equal mixture of
electrons and positrons, the latter will be much more ``mobile'', leading
to comparable currents in the two species and to a larger total
current. In addition, because in our simulations the magnetization
$\sigma$ is kept constant, the initial (and subsequent) magnetic field is
smaller for the $e_{-}$-$e_{+}$ mixture (see Tab.~\ref{TableI}). Using
Eq.~\eqref{eq:RXP_Ez}, $\mathcal{R}_{_{\rm XP}} \sim E^z / B_{xy} \sim
J_{xy} B_{xy} / (e\,n\, B_{xy}) \sim (\nabla \times B)_{xy}/(e\,n) \sim
B_{0,z} / (e \, n\, \mathcal{L})$, where $\mathcal{L}$ is a
characteristic lengthscale for the current sheet. It follows that
$\mathcal{R}_{_{\rm XP}}|_{\chi \to 0}/\mathcal{R}_{_{\rm XP}}|_{\chi \to
  1} \sim B_{0,z} |_{\chi \to 1} / B_{0,z}|_{\chi \to 0} \sim 3$ (see
Tab.~\ref{TableI}) if we assume that the thickness of the current sheet
$\mathcal{L}$ is not very different in the two cases. Such a ratio of the
reconnection rate that we have estimated in this way is similar to what
is shown in Fig.~\ref{Image_2}. Stated differently, in general we
  expect that the reconnection rate will decrease with the increase of
  $\chi$ as the addition of positrons will ``pollute'' any plasma
  mixture. As a result, in the electron-positron mixture relative to
$\chi=0.9$, the number of X-points will increase, but because of the
reduced magnetic field they will have a smaller reconnection rate and be
responsible for a smaller dissipation. 

Although our results offer an approach to studying and understanding
relativistic multi-species plasmas, several improvements are
possible. First, extending the simulations to three spatial dimensions
would allow us to assess how the plasmoid phenomenology, which is known
to change in three dimensions~\citep[see, e.g.,][]{NathanailEA22},
affects the conclusions drawn here. Second, it would be valuable to
determine how high-curvature spacetimes influence the inertial effects of
charged particles. Finally, while the spatial and particle resolutions
employed in this study are sufficient for drawing robust conclusions,
higher resolutions are necessary to further explore the spectral
properties of the turbulent plasma. We plan to address these aspects in
future work.

\section*{Acknowledgements}

It is a pleasure to thank F. Camilloni for useful discussions. This
research is supported from the European Union's Horizon Europe research
and innovation programme under grant agreement No. 101082633 (ASAP), the
ERC Advanced Grant ``JETSET: Launching, propagation and emission of
relativistic jets from binary mergers and across mass scales'' (Grant
No. 884631), and the ERC Consolidator Grant SPAWN (Grant
No. 863412). A.C.O acknowledges to the DGAPA-UNAM (grant IN110522), the
``Ciencia Básica y de Frontera 2023–2024'' program of SECIHTI México
(projects CBF2023-2024-1102 and 257435). L.R. acknowledges the Walter
Greiner Gesellschaft zur F\"orderung der physikalischen
Grundlagenforschung e.V. through the Carl W. Fueck Laureatus Chair and
the hospitality at CERN, where part of this research was carried
out. Computational resources were provided by CINECA through the ISCRA
Class B project ``KITCOM-HP10BB7U73''. We acknowledge ISCRA for awarding
this project access to the LEONARDO supercomputer, owned by the EuroHPC
Joint Undertaking, hosted by CINECA (Italy). Simulations were also
performed on HPE Apollo HAWK at the High Performance Computing Center
Stuttgart (HLRS) under the grant BNSMIC, on the Goethe-HLR supercomputer,
on the local supercomputing cluster Calea, and on the Alarico HPPC
Computing Facility at the University of Calabria.


\appendix
\renewcommand{\thefigure}{A\arabic{figure}}
\setcounter{figure}{0}

\section{Consistency Study}
\label{sec:consistency}

As discussed in the main text, choosing a resolution of $\Delta x = 4/3
\, d_{e^-}$ does not fully resolve the electron skin depth. To verify
that this choice does not compromise the validity of our results, we
perform a convergence study of the particle distribution functions across
multiple configurations with varying grid spacings, while keeping the
physical box length fixed at $L_0 = 5461.33 \, d_{-}$. Specifically, we
analyze the following resolutions
\begin{eqnarray}
  &\Delta x = (1/3)\, d_{e^-}\,, & N_x = N_y = 16384\,, ~~ \text{PPC} =
  15\,,\nonumber \\
  &\Delta x = (2/3)\, d_{e^-}\,, & N_x = N_y = 8192\,, \quad \text{PPC} =
  60\,,\nonumber \\
  &\Delta x = (4/3)\, d_{e^-}\,, & N_x = N_y = 4096\,, \quad \text{PPC} =
  240\,,\nonumber \\
  &\Delta x = (8/3)\, d_{e^-}\,, & N_x = N_y = 2048\,, \quad \text{PPC} =
  960\,,\nonumber \\
  &\Delta x = (16/3)\, d_{e^-}\,, & N_x = N_y = 1024\,, \quad \text{PPC} =
  3840\,. \nonumber \\
  \label{eq:num_setups}
\end{eqnarray}
where $\texttt{PPC}$ indicates the number of particles per cell.

\begin{figure}
  \centering	
  \includegraphics[width=1.0\columnwidth]{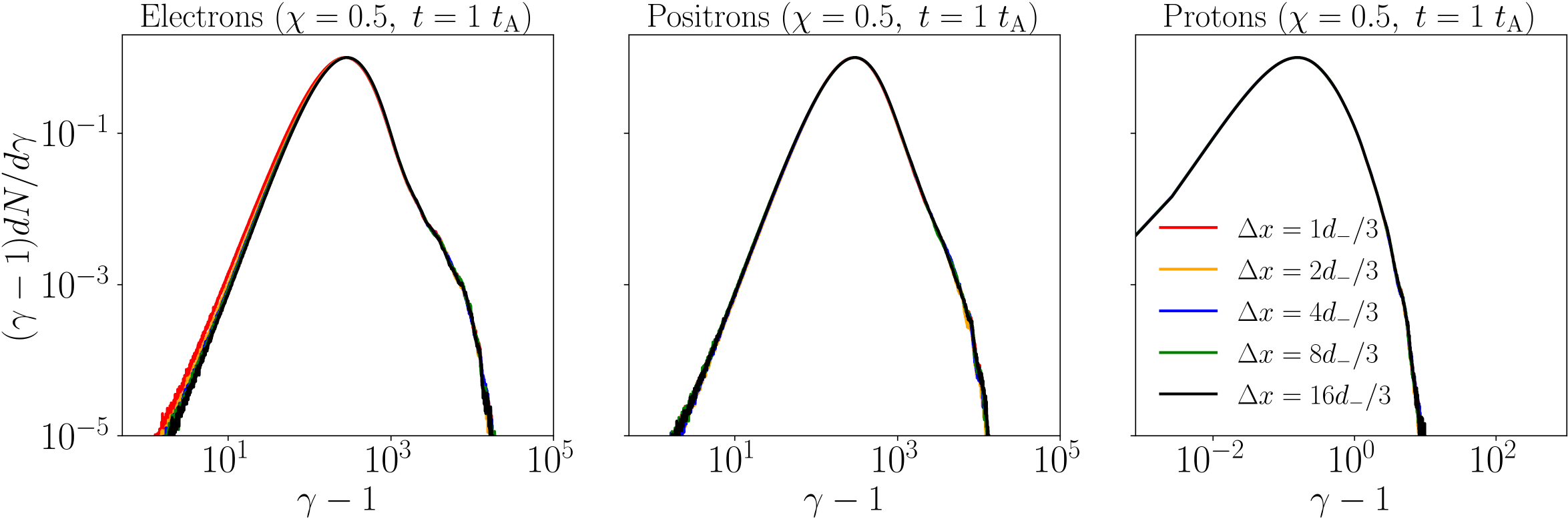}
  \caption{Consistency study showing the particle kinetic-energy
    distributions expressed as a function of the Lorentz factor for
    electrons (left panel), positrons (middle panel), and protons (right
    panel). The simulations refer to the case with $\chi = 0.5$ and show
    the different setups listed in~\eqref{eq:num_setups} at $t =
    t_{\text{A}}$}.
  \label{Image_apx1}
\end{figure}

Concentrating for convenience on the case with $\chi = 0.5$, the results
of this consistency analysis are summarized in Fig.~\ref{Image_apx1},
which reports the particle kinetic-energy distributions (with the
rest-mass energy subtracted) expressed as function of the Lorentz
factor for electrons (left panel), positrons (middle panel), and protons
(right panel), along with the numerical setups detailed
in~\eqref{eq:num_setups}. Clearly, all distributions are very similar to
each other, especially at the highest energies. This confirms that a
resolution of $\Delta x = (4/3) \, d_{e^-}$ is sufficient for our
purposes, as it does not violate the main PIC constraints outlined in the
core of the paper and provides a sufficient number of PPC to contrast and
limit noise.

\begin{figure}
	\centering
	\includegraphics[width=1.0\columnwidth]{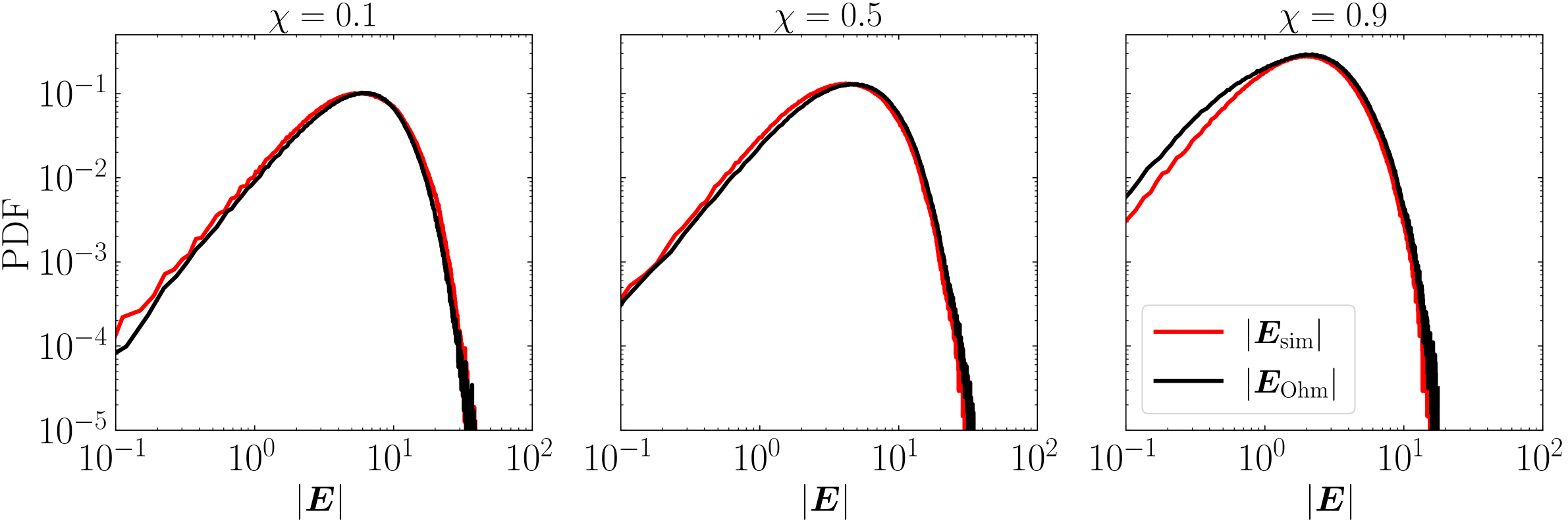}
	\caption{PDFs of the modulus of the electric field $|\bm{E}|$ at time
		$t = 3\,t_\text{A}$ as computed by the aggregation of the various
		components ($|\bm{E}_{\rm Ohm}|$, black line) or by the total
		electric field of the simulation ($|\bm{E}_{\rm sim}|$, red
		line). The different panels, refer to $\chi = 0.1$ (left), $\chi =
		0.5$ (middle) $\chi = 0.9$ (right).}
	\label{Image_2_appendix}
\end{figure}

Another important consistency check in our simulations is provided by
combining together all the different electric-field components in the
last line of Ohm's law in Eq.~\eqref{eq:Ohm}, and that we compute
separately, and to ensure that they actually lead to the same total
electric field computed in the simulation. This is shown in
Fig.~\ref{Image_2_appendix}, which reports the PDFs of the modulus of the
electric field $|\bm{E}|$ at time $t = 3\,t_\text{A}$ as computed by the
aggregation of the various components ($|\bm{E}_{\rm Ohm}|$, black line)
or by the total electric field of the simulation ($|\bm{E}_{\rm sim}|$,
red line). The different panels, which refer to $\chi = 0.1$ (left),
$\chi = 0.5$ (middle) $\chi = 0.9$ (right) clearly show that the
different contributions reconstruct very accurately the total electric
field; small differences appear for the case with $\chi = 0.9$ but only
in the weakest fields, while the larger ones are reproduced very
accurately. We expect this deviation to be linked to the electric-field
contribution $\bm{E}_{\partial t} \propto \partial_t \bm{\mathcal{J}}$,
which is not well-captured on the largest scales as the number of positron
increases with $\chi$~\citep[see also][]{hesse2007dissipation} Overall,
the results in Fig.~\ref{Image_2_appendix} provide us with important
confidence on our ability to track down the single contributions to Ohm's
law.

\section{On the Generalized Ohm's Law}
\label{AppA}

We here provide additional details on the derivation of the generalized
Ohm's law of a three-species relativistic plasma, Eq.~\eqref{eq:Ohm}
reported in the main text. As customary, we start from the relativistic
Vlasov equation, governing for each species $a$ the evolution of the
corresponding distribution function $f_a$ and taking the form [see
Chap. 2 in~\citet{cercignani2002relativistic} or Chap. 6
in~\citet{liboff2003kinetic}]
\begin{equation}
	\partial_t f_a + \frac{\bm{u}}{\gamma} \cdot \bm{\nabla} f_a + \frac{q_a}{m_a}
	\left( \bm{E} + \frac{\bm{u} \times \bm{B}}{\gamma} \right) \cdot
	\frac{\partial f_a}{\partial \bm{u}} = 0\,,
	\label{Vlas}
\end{equation}
where $f_a = f_a(\bm{x}, \bm{u}, t)$ is the particle distribution
function, $\bm{u} = \gamma \bm{v}$ the four velocity, with $\bm{v}$ the
three-velocity and $\gamma := (1 - v^i v_i)^{-1/2}$ the corresponding
Lorentz factor (for simplicity, we omit the index in the velocities).

The moment equations are obtained by multiplying Eq.~\eqref{Vlas} by
powers of the four-velocity $\bm{u}$ and integrating over
$\bm{u}$-space. In particular, the first moment of the relativistic
Vlasov equation yields the continuity equation
\begin{equation}
	\!\!\!\!\! \partial_t \int d^3 u f_{a} + \bm{\nabla} \cdot \int d^3 u
	\frac{\bm{u}}{\gamma} f_{a} = \partial_t n_{a} + \bm{\nabla} \cdot (n_{a}
	\bm{V}_{a}) = 0\,,
\end{equation}
while the second moment provides the momentum equation
\begin{equation}
	\!\!\!\!\! m_{a} n_{a} \left( \partial_t \bm{U}_{a} \!\! + \!\!
	\bm{V}_{a} \cdot \bm{\nabla} \bm{U}_{a} \right) + \bm{\nabla} \cdot
	\bm{{P}}_{a} = q_{a} n_{a} \left( \bm{E} + \bm{V}_{a}
	\times \bm{B} \right)\,,
	\label{Ohm1}
\end{equation}
where the pressure tensor $\bm{{P}}_{a}$ is defined solely in terms of
thermal inertia effects
\begin{equation}
	\begin{aligned}
		\bm{{P}}_{a} & := \int d^3u \, m_{a} \frac{\bm{u}
			\bm{u}}{\gamma} f_{a} - m_{a} n_{a} \bm{V}_{a} \bm{U}_{a} \\[2pt] & =
		\bm{{\bm{\Pi}}}_{a} - m_{a} n_{a} \bm{V}_{a}
		\bm{U}_{a}\,.
		\label{eq:press_tensor_a}
	\end{aligned}
\end{equation}

Multiplying the continuity equation by $m_{a} \bm{U}_{a}$ and
substituting into Eq.~\eqref{Ohm1} after using the vector identity $n_{a}
\bm{V}_{a} \cdot \bm{\nabla} \bm{U}_{a} = \bm{\nabla} \cdot (n_{a}
\bm{V}_{a} \bm{U}_{a}) - \bm{U}_{a} \bm{\nabla} \cdot (n_{a} \bm{V}_{a})$
yields a more convenient form
\begin{equation}
	\begin{aligned}
		& m_{a} \left[ \frac{\partial}{\partial t} (n_{a} \bm{U}_{a}) +
		\bm{\nabla} \cdot (n_{a} \bm{V}_{a} \bm{U}_{a}) \right] + \nabla
		\cdot \bm{{P}}_{a} - \\ & q_{a} n_{a} (\bm{E} + \bm{V}_{a}
		\times \bm{B}) = 0\,.
		\label{eq:mom_eq_comp}
	\end{aligned}
\end{equation}
Noting that the inertial term in the first line of
Eq.~\eqref{eq:mom_eq_comp} cancels with the contribution from the
pressure tensor~\eqref{eq:press_tensor_a}, we obtain the following, more
compact form of the momentum equation or, equivalently, of Ohm's law
\begin{equation}
	m_{a} \frac{\partial}{\partial t} (n_{a} \bm{U}_{a}) + \bm{\nabla} \cdot
	\bm{{\bm{\Pi}}}_{a} - q_{a} n_{a} (\bm{E} + \bm{V}_{a} \times \bm{B}) =
	0\,,
\end{equation}
where the distinction between inertial and thermal effects is lost.

To derive the expression for the electric field, we apply this
formulation to all plasma species by multiplying each equation by
$q_{a}/m_{a}$ so as to obtain the following set of equations
\begin{equation}
	\begin{aligned}
		& \frac{\partial}{\partial t} (e n_{+} \bm{U}_{+}) + \frac{e}{m_-}
		\bm{\nabla} \cdot \bm{{\bm{\Pi}}}_{+} -
		\frac{e^2 n_{+}}{m_-} (\bm{E} + \bm{V}_{+} \times \bm{B}) = 0\,, \\ &
		\frac{\partial}{\partial t} (e n_{-} \bm{U}_{-}) + \frac{e}{m_-}
		\bm{\nabla} \cdot \bm{{\bm{\Pi}}}_{-} +
		\frac{e^2 n_{-}}{m_-} (\bm{E} + \bm{V}_{-} \times \bm{B}) = 0\,, \\ &
		\frac{\partial}{\partial t} (e n_p \, \bm{U}_p) + \frac{e}{m_p}
		\bm{\nabla} \cdot \bm{{\bm{\Pi}}}_p -
		\frac{e^2 n_p}{m_p} (\bm{E} + \bm{V}_p \times \bm{B}) = 0\,. \\
	\end{aligned}
\end{equation}
Summing them to obtain the displacement momentum current density and
solving for the electric field yields
\begin{equation}
	\begin{aligned}
		\bm{E} & = \frac{m_-}{e^2 \mathcal{N}} \, \partial_t \bm{\mathcal{J}} -
		\frac{1}{\mathcal{N}} \left( n_{+} \bm{V}_{+} + n_{-} \bm{V}_{-} \right)
		\times \bm{B} \\ & \phantom{=} - \frac{1}{\mathcal{N}} \, n_p \bm{V}_p
		\times \bm{B} - \frac{1}{e \mathcal{N}} \, \bm{\nabla} \cdot
		\bm{{\bm{\Pi}}}_{-} + \frac{1}{e \mathcal{N}} \, \bm{\nabla} \cdot
		\bm{{\bm{\Pi}}}_{+} \\ & \phantom{=} + \frac{m_-}{m_p} \frac{1}{e
			\mathcal{N}} \, \bm{\nabla} \cdot \bm{{\bm{\Pi}}}_p\,,
		\label{eq:Ohm_app}
	\end{aligned}
\end{equation}
where we recall that $\mathcal{N} := n_p {m_-}/{m_p} + n_{+} + n_{-}$ and
$\bm{\mathcal{J}} := \sum_{a} q_{a} n_{a}
\bm{U}_{a}$. Expression~\eqref{eq:Ohm_app} clearly coincides
Eq.~\eqref{eq:Ohm} in the main text.

\section{Appendix C: Additional Statistics}
\label{AppB} 

\begin{figure}
	\centering
	\includegraphics[width=0.99\columnwidth]{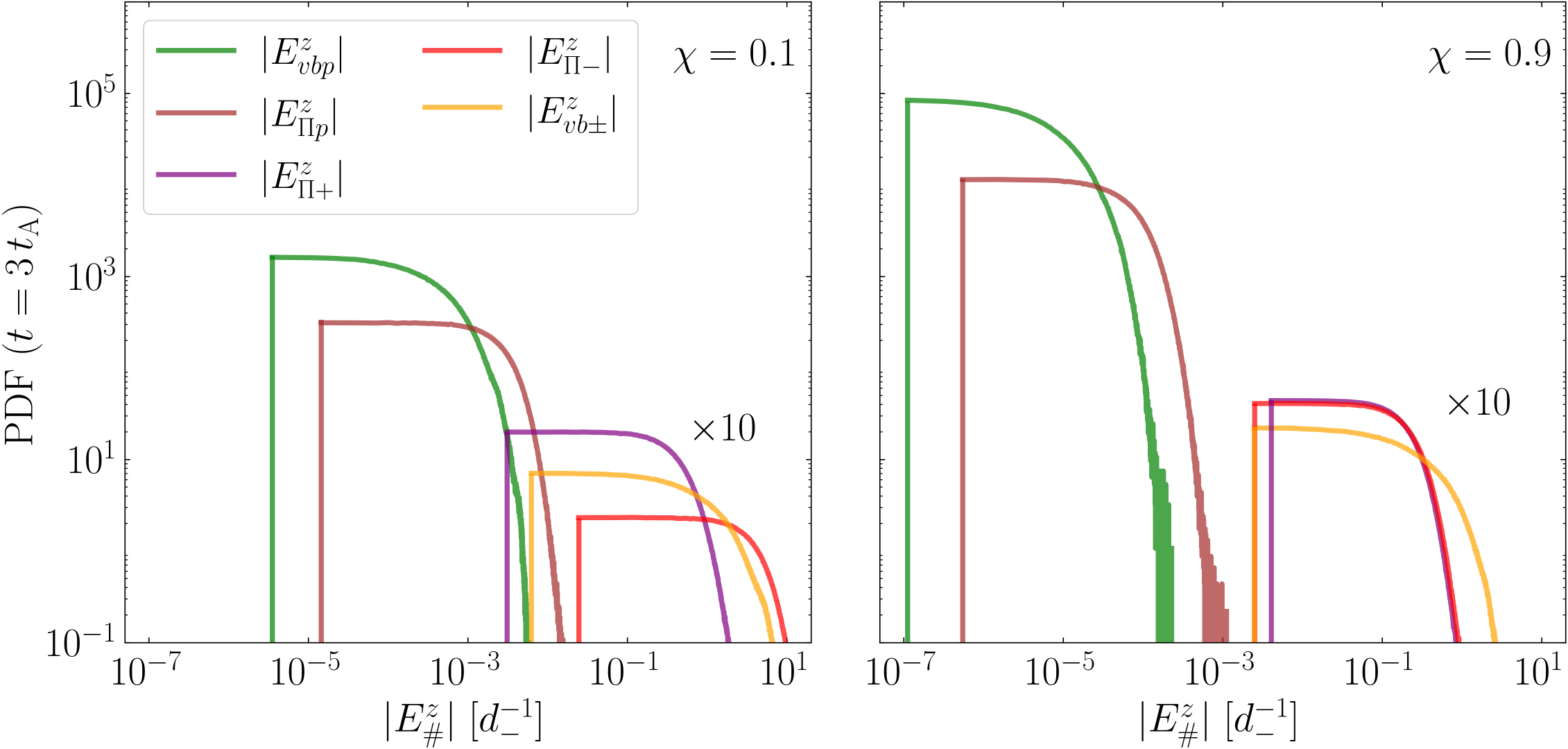}
	\caption{The same as in Fig.~\ref{Image_2} but for $\chi=0.1$ (left)
		and $\chi=0.9$ (right). Much of what discussed for the case
		$\chi=0.5$ in Fig.~\ref{Image_2} applies also for the other
		concentrations.}
	\label{Image_3_appendix}
\end{figure}

Finally, we present in Fig.~\ref{Image_3_appendix} information
complementary to that in Fig.~\ref{Image_2} of the main text, and report
the PDF of the various terms contributing to the $z$ component of the
electric field at $t = 3 \,t_{\text{A}}$, computed over the whole
domain. In contrast to Fig.~\ref{Image_2}, which was computed for
$\chi=0.5$, the panels in Fig.~\ref{Image_3_appendix} refer to plasmas
with $\chi=0.1$ (left) and $\chi=0.9$ (right), respectively.

A rapid comparison among the different cases highlights a behavior we
have already emphasized. The almost pure electron-proton plasma ($\chi =
0.1$) produces sizeable contributions to the PDFs of the electromotive and
pressure-gradient components of the electric field, i.e., $E^z_{vbp}$ and
$E^z_{\Pi p}$, respectively. Their strength, however, is subdominant and
this is also true when considering an almost pure electron-positron
plasma ($\chi = 0.9$). Hence, for all values of $\chi$, the dominant
contributions to the electric field come from the electron-positron
electromotive component $E^z_{vb \pm}$ and the pressure-gradient terms
$E^z_{\Pi +}$ and $E^z_{\Pi -}$. The latter satisfy $E^z_{\Pi -} \gtrsim
E^z_{\Pi +}$ for $\chi =0.1$ and $E^z_{\Pi -} \sim E^z_{\Pi +}$ for $\chi
=0.9$, as expected.

Finally, we comment on another widely used method to compute the
reconnection rate at each X-point is to employ the plasma drift velocity
in the vicinity of the reconnection site~\citep[see,
e.g.,][]{MellahEA22}, which here we model as
\begin{equation}
	\langle \mathcal{R}_{\text{XP}} \rangle := \langle
	|\bm{E}\times\bm{B}|_{\hat{n}}/(B^2_{\hat{n}} + B^2_{\hat{t}}) \rangle_{\text{XP}}\,.
	\label{eq:RXP_vd}
\end{equation}

Ideally, our expectation was that the reconnection rate computed via
Eq.~\eqref{eq:RXP_vd} would yield a statistics similar to that
encountered with Eq.~\eqref{eq:RXP_vd} and shown in
Fig.~\ref{Image_2_appendix}. In reality, we found that besides being less
robust, the use of the drift velocity yields almost uniform PDFs of
$\langle \mathcal{R}_{\text{XP}} \rangle$, thus making it difficult to
determine a median value. This results suggests that the use of the drift
velocity may be useful in highly magnetized plasmas, i.e., with $\sigma
\gg 1$, where the motion of particles is constrained to follow the
magnetic-field lines and the dynamics can be well approximated by the
frozen-in field regime. Under these conditions, the drift velocity
represents the transport velocity of the magnetic field lines and can
provide a local estimate of the reconnection rate. However, in scenarios
like the one explored here, where $\sigma \approx 1$ and turbulence is
fully developed, the drift velocity also acquires a stochastic nature.
Plasma inertia is comparable to, or even exceeds, the magnetic
tension forces, and the local dynamics is dominated by effects such as
compression, misalignments between $\bm{E}$ and $\bm{B}$, instabilities
(e.g., drift-kink), and pressure anisotropies. In this context, the
quantity $\bm{E} \times \bm{B} / B^2$ can increase significantly near
local minima of $B^2$, as often occurs at X-points, leading to physically
unreliable estimates of the reconnection rate. Overall, these results
suggest that while reconnection rates may and should be estimated via
multiple different methods, not all of them are equally adequate.



\providecommand{\noopsort}[1]{}\providecommand{\singleletter}[1]{#1}%


\end{document}